\documentclass[useAMS,usenatbib,graphicx]{mn2e}

\usepackage{graphicx}
\usepackage{lineno}

\title[Sheets, filaments and clumps]{Sheets, filaments and clumps - high resolution simulations of how the thermal instability can form molecular clouds}
\author[C. J. Wareing et al.]{C. J. Wareing$^{1}$\thanks{E-mail:
C.J.Wareing@leeds.ac.uk}, S. A. E. G. Falle$^{2}$ and J. M. Pittard$^{1}$\\
$^{1}$School of Physics and Astronomy, University of Leeds, Leeds, LS2 9JT, U.K.\\
$^{2}$School of Mathematics, University of Leeds, Leeds, LS2 9JT, U.K.}
\begin{document}

\date{Accepted 2019 March 6. Received 2019 February 19; in original form 2018 December 21}

\pagerange{\pageref{firstpage}--\pageref{lastpage}} \pubyear{2002}

\maketitle

\label{firstpage}

\begin{abstract}
This paper describes 3D simulations of the formation of collapsing cold clumps via 
thermal instability inside a larger cloud complex. The initial condition was a diffuse 
atomic, stationary, thermally unstable, 200\,pc diameter spherical cloud in pressure 
equilibrium with low density surroundings. This was seeded with 10\% density
perturbations at the finest initial grid level (0.29\,pc) around n$_{\rm H}$=1.1\,cm$^{-3}$
and evolved with self-gravity included. No magnetic field was imposed. Resimulations  
at a higher resolution of a region extracted from this simulation (down to 0.039\,pc),  
show that the thermal instability forms sheets, then filaments and finally clumps.
The width of the filaments increases over time, in one particular case from 0.26 
to 0.56\,pc. Thereafter clumps with sizes of around 5\,pc grow at the intersections 
of filaments. 21 distinct clumps, with properties similar to those observed in molecular 
clouds, are found by using the FellWalker algorithm to find minima in the gravitational
potential. Not all of these are gravitationally bound, but the convergent nature of  
the flow and increasing central density suggest they are likely to form stars. Further 
simulation of the most massive clump shows the gravitational collapse to a density  
$> 10^6$\,cm$^{-3}$. These results provide realistic initial conditions that can be 
used to study feedback in individual clumps, interacting clumps and the entire 
molecular cloud complex.
\end{abstract}


\section{Introduction}

Extensive studies of the nearest star-forming clouds, most recently with the {\it Herschel} Space
Observatory have revealed that every interstellar cloud contains an intricate network of interconnecting
filamentary structures \cite[see, for example, Section 2 of the review of][ and references therein]{andre14}. 
The data, from {\it Herschel} and near-IR studies for example, suggest a scenario 
in which these ubiquitous filaments represent a key step in the star formation process: large-scale flows
compress the diffuse ISM and form molecular clouds; an interconnecting filamentary structure forms
within these clouds; magnetic fields affect the flow of material and hence overall structure, although 
do not appear to set the central densities in the filaments; gravity plays an increasingly important role, 
fragmenting the filaments once they are cold and dense into prestellar cores and finally protostars, commonly
occuring at the intersections or hubs of the intricate structure. 

Observational results now connect well with numerical simulations, as highlighted in Section 5 
of \cite{andre14} and references therein. Early numerical simulations showed that
gas is rapidly compressed into a hierarchy of sheets and filaments, without the aid of gravity
\citep*{bastien83,porter94,vazquez94,padoan01}. Turbulent box simulations and colliding flows produce filaments
\citep[e.g.][]{maclow04,hennebelle08,federrath10,gomez14,moeckel15,smith14,kirk15}.
\cite{hennebelle13} demonstrated the formation of filaments through the velocity shear that
is common in magnetised turbulent media. Other authors have explained filaments as the stagnation
regions in turbulent media \citep{padoan01}. The formation of filaments preferentially perpendicular 
to the magnetic field lines is possible in strongly magnetised clouds \citep{li10,wareing16}.
\cite{andre14} note that the same 0.1\,pc filament width is measured for low-density, subcritical
filaments suggesting that this characteristic scale is set by the physical processes producing the 
filamentary structure. Furthermore, they note that at least in the case of diffuse gravitationally
unbound clouds (e.g. Polaris), gravity is unlikely to be involved. Large-scale compression
flows, turbulent or otherwise, provide a potential mechanism, but it is not clear why any of these
would produce filaments with a constant radius. 

\cite{smith14} examined the influence of different types of turbulence, keeping the initial mean density
constant in simulations without magnetic fields. They found that when fitted with a Plummer-like profile, the
simulated filaments are in excellent agreement with observations, with shallow
power-law profile indexes of p $\approx$ 2.2, without
the need for magnetic support. They found an average FWHM of $\approx$ 0.3\,pc, 
when considering regions up to 1\,pc from the filament centre, in agreement with predictions for
accreting filaments. Constructing the fit using only the inner regions, as in {\it Herschel} observations,
they found a resulting FWHM of $\approx$ 0.2\,pc.

\cite{kirk15} used the FLASH hydrodynamics code to perform numerical simulations of turbulent cluster-forming 
regions, varying density and magnetic field. They used HD and MHD simulations, initialised with a supersonic
($M \approx 6$) and super-Alfv{\'e}nic ($M_A \approx 2$) turbulent velocity field, chosen to match 
observations, and identified filaments in the resulting column density maps. They found magnetic
fields have a strong influence on the filamentary structure, tending to produce wider, less centrally peaked and 
more slowly evolving filaments than in the hydrodynamic case. They also found the magnetic field  is able to
suppress the fragmentation of cores, perhaps somewhat surprisingly with super-Alfv{\'e}nic motion
involved in the initial condition. Overall, they noted the filaments formed in their simulations have properties 
consistent with the observations they set out to reproduce, in terms of radial column density profile,
central density and inner flat radius.

Numerical simulations now include the thermodynamic behaviour of the cloud material, magnetic fields, gravity
and feedback from massive stars, both radiative and dynamic 
\citep{beuther08,harper09,krumholz09,gray11,koenig12,rogers13,rogers14,gatto15,offner15,walch15,walch15b,giri15,kortgen16}. 
Supersonic, trans-Alfv{\'e}nic turbulence 
has emerged as an ingredient which can, when injected at the right scale, result in the formation of 
filaments which possess properties remarkably similar to those derived from observational results
whilst also reproducing the observed relatively smooth nature of the magnetic field. 
The interested reader is referred to lengthy introductions and reviews in the first sections
of previous works by the same authors \citep{wareing16,wareing17a,wareing17b,wareing18}.

The work presented here continues the exploration of the formation of clumps connected 
by filamentary structures through the use of hydrodynamic simulations of the thermal instability 
\citep{parker53,field65}. A number of authors have investigated analytically the effects of 
different mechanisms on the thermal instability \citep*{birk00,nejad03,stiele06,fukue07,shadmehri09}.
Other groups have numerically investigated flow-driven molecular cloud formation including the effects of the 
thermal instability \citep*[e.g.][]{lim05,vazquez07,hennebelle08,heitsch09,ostriker10,vanloo10,inoue12}.
\citet{hennebelle07a} and \citet{hennebelle07b} showed that H{\sc i} clouds formed by thermal instability
can explain a variety of observational characteristics. \citet{inoue12} used 3D MHD simulations, including
radiative cooling and heating, to investigate the formation of molecular clouds. They consider
the scenario of accretion of H{\sc i} clouds and the piling up of the initial H{\sc i} medium behind shock
waves induced by accretion flows in order to form a molecular cloud. They find the resulting
timescale of molecular cloud formation of $\sim$10\,Myr is consistent with the evolutionary
timescale of the molecular clouds in the LMC \citep{kawamura09}. 
This numerical work has included magnetic fields, self-gravity and the thermal instability and has identified the 
thermal and dynamical instabilities that are responsible for the rapid fragmentation
of the nascent cloud, largely through flow-driven scenarios.

Here we concentrate on the thermal instability itself without any initial flow in a low-density cloud of quiescent 
diffuse medium initially in the unstable phase. The cloud is in
pressure equilibrium with its lower-density (thus temperature above equilibrium) surroundings and we 
include accurate thermodynamics and self-gravity. In particular, the aim is to discern whether
thermal instability alone can create high enough density structure that gravity can then
dominate and drive the eventual collapse of the clump to form clusters of stars,
regardless of the apparent status of the energy budget. This has not been observed
in our previous works. Subsequent aims are to confirm 
whether such clumps display realistic properties when compared to
observationally-derived properties, and whether mergers and collisions of clumps 
reproduce observations.
The simulations presented herein are also intended as a means to define
realistic initial conditions for cluster feedback simulations.

In the next section the initial condition for this high resolution study is described. In
Section \ref{methods}, the numerical method and model are discussed. 
Results are discussed in the following sections.
Specifically, the evolution of the whole cloud is discussed in Section \ref{evolve}.
The state of the cloud complex and the clumps identified within are discussed in
Section \ref{snapshot}. Density and velocity power spectra of the cloud at this time 
and an individual clump are presented in Section \ref{powspec} and the final
gravitational collapse of the most massive clump is discussed in Section \ref{collapse}.
The work is summarised and concluded in Section \ref{conclusions}, which also sets 
out how details of the simulations may be used in future works.

\section{Initial conditions}\label{inits}

\begin{figure*}
\centering
\includegraphics[width=160mm]{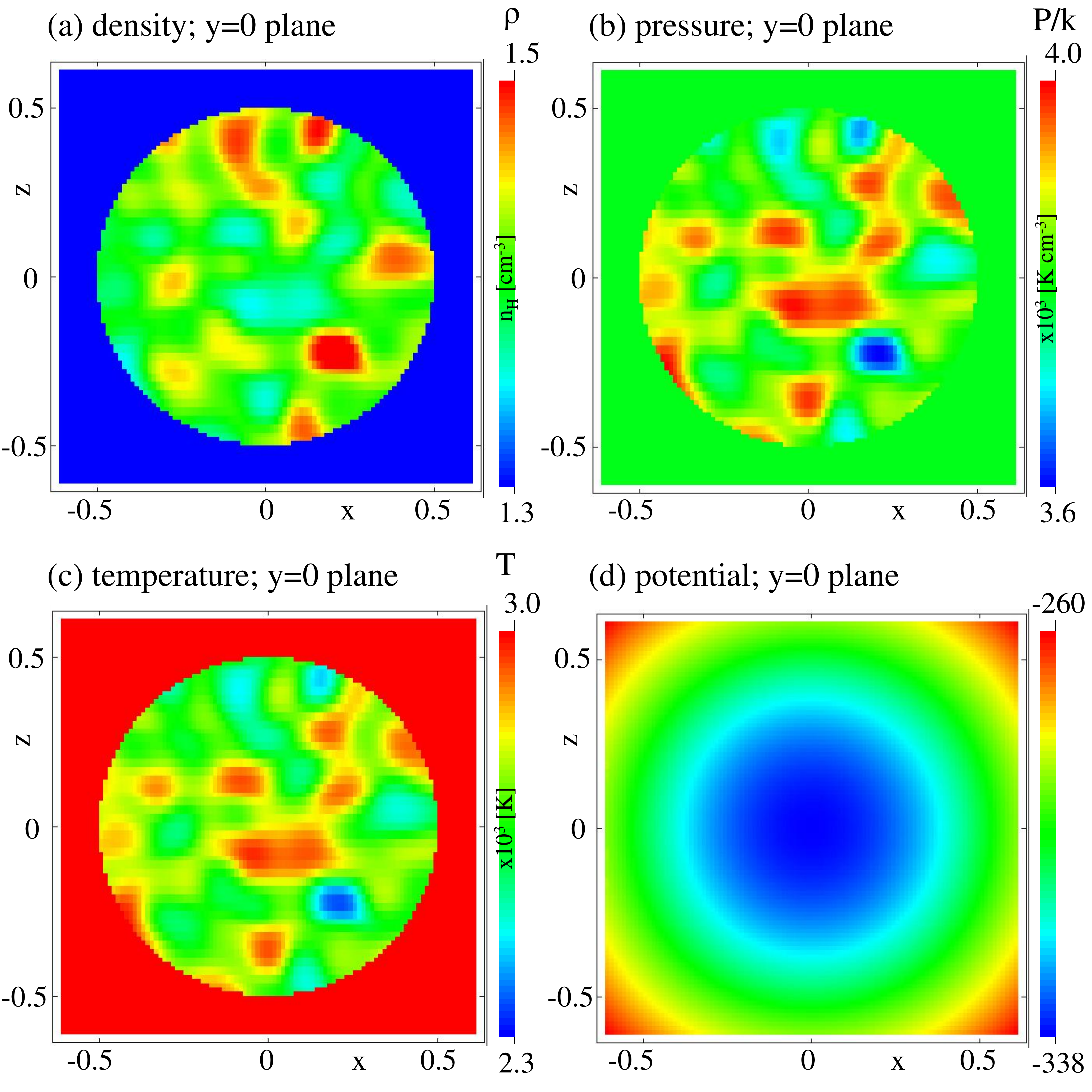}
\caption{Initial conditions in the Model 1 simulation. 
The appearance of the extracted region of the Paper IV HD simulation on the $y=0$ plane, surrounded 
by the ambient pressure-matched stable conditions, imposed
in the Model 1 simulation. 
The region was extracted at t=16.2\,Myrs of the Paper IV HD simulation.
The unit of distance is 50\,pc.
Raw data is available from https://doi.org/10.5518/483 .}
\label{initial}
\end{figure*}

Our recent work \citep{wareing16,wareing17a,wareing17b,wareing18} 
has highlighted the way that the thermal instability \citep{field65}, 
under the influence of gravity and realistic magnetic fields (magnetic pressure)
in the ideal magnetohydrodynamic (MHD) limit, 
can drive the evolution of diffuse thermally unstable warm clouds from a 
pressure-supported quiescent low-density state to form high-density, cold ($\leq$100\,K)
clumps and, in the presence of magnetic field, sheet-like structures that are filamentary 
in appearance, perpendicular to the applied field \citep[hereafter Paper I]{wareing16}.

In our most recent work \cite[hereafter Paper IV]{wareing18}, the
model of stellar wind feedback in a sheet-like structure first
considered in \citet[hereafter Paper II]{wareing17a} was applied
to the Rosette Nebula, solving outstanding issues 
around the age and size of the nebula \citep{bruhweiler10}.
The calculations in Paper IV explored both hydrodynamic (HD) and MHD initial 
conditions. The computational domain extended from -150\,pc to +150\,pc
in all three Cartesian $xyz$ directions. A diffuse cloud with an average number density of atomic hydrogen 
of n$_H$ = 1.1 cm$^{-3}$ and radius r=100\,pc was
placed at the centre of the domain (0,0,0). The cloud was seeded with random 
density variations of 10\% about its average density. The cloud mass was 
135,000\,M$_{\odot}$. Initial pressure was set according to the (unstable) equilibrium 
of heating and cooling at P$_{eq}$/k = $4700\pm300$ K\,cm$^{-3}$, resulting
in an initial cloud temperature T$_{eq}$ = $4300\pm700$\,K. External to the cloud the 
density was reduced by a factor of 10 to n$_H$ = 0.1 cm$^{-3}$, but the
external pressure matched the cloud (P$_{eq}$/k = $4700$
K\,cm$^{-3}$). The external medium was prevented from cooling or
heating, keeping this pressure throughout the simulation.
No magnetic field or velocity structure was introduced in the HD
case. When the simulation was evolved, condensations 
due to the thermal instability began to grow in the cloud and after 16\,Myrs 
their densities were $\sim 40$ per cent greater than
the initial average density of the cloud.

\begin{table*}
\begin{center}
  \caption{Suite of 3D Cartesian hydrodynamic simulations with self-gravity performed in this work. Details
of the HD Simulation in Paper IV, from which the central region is extracted for resimulation in this work, are 
provided for reference.}
  \label{table-sims}
  \begin{tabular*}{176mm}{lllllllllll}
\hline
Name & Domain & G0 & Levels & t$_{start}$ & Resolution & HD &Cooling?& Heating? & Gravity?\\
& pc on a side& \# of cells& of AMR & [Myr] & [pc] & /MHD?\\
\hline
Paper IV HD & 300 & $8\times8\times8$ & 8 & 0.0 & 0.29 &HD & Y & Y & Y\\
\citep{wareing18} \\
\hline
Model 1 & 100 & $10 \times 10 \times 10$ & 6 & 16.2 & 0.3125 &HD & Y & Y & Y\\
Model 2 & 100 & $10 \times 10 \times 10$ & 7 & 21.7 & 0.156 &HD & Y & Y & Y\\
Model 3 & 100 & $10 \times 10 \times 10$ & 8 & 21.7 & 0.078 &HD & Y & Y & Y\\
Model 4 & 100 & $10 \times 10 \times 10$ & 9 & 22.8 & 0.039 &HD & Y & Y & Y\\
\hline
Clump & 10 & $10 \times 10 \times 10$ & 7 & 44 & 0.0156 &HD & Y & Y & Y\\
\hline
  \end{tabular*}
\end{center}
\end{table*}

In the present work, the central region of Paper IV Simulation 1, the 
HD simulation, has been extracted at a simulation time of 
16\,Myrs, placed in ambient pressure-matched surroundings and then evolved at higher 
resolution in order to study the cloud complex formed and to determine the nature 
of the clumps therein. The aim is to
explore how gravity takes over from the effect of thermal instability, without the
extra complexity of magnetic field at this stage. If thermal instability, subsequently dominated
by gravity is able to create truly star-forming collapsing clumps with realistic properties, then a 
particularly simple scenario for the formation of stars is presented.
Simulations in a future work will examine the role of magnetic field in the formation
of cloud complexes and collapsing molecular substructure.
The average pressure across the cloud in the Paper IV HD simulation had decreased from the initial pressure 
to levels around P/k = $3800\pm200$ K\,cm$^{-3}$. In the Paper IV HD 
simulation, this decrease in cloud pressure compared to the
over-pressured surroundings resulted in a compression of the cloud by the surrounding medium.
Isolating the study of the thermal instability from this effect is the secondary reason, 
after allowing for higher resolution, for extracting the 25\,pc-radius sphere centred on (0,0,0)
in this new simulation.

\begin{figure}
\centering
\includegraphics[width=85mm]{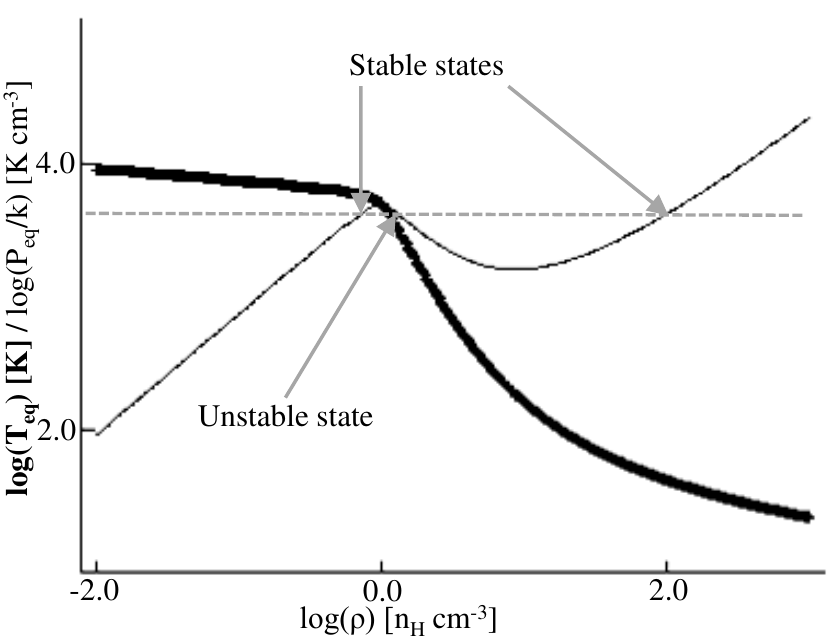}
\caption{Equilibrium curves. Thermal equilibrium pressure ($P_{eq}/k$ - thin line) 
and temperature ($T_{eq}$ - thick line) vs. density for the cooling and heating 
functions selected in this work. Where the grey dashed line intersects the 
pressure curve (thin black line)
indicates, from left to right, the stable, unstable and stable equilibrium states.
Raw data is available from https://doi.org/10.5518/483 .}
\label{eqm}
\end{figure}

A slice through the centre of the new domain, containing the extracted section 
within r~$\leq$~0.5 (25\,pc) is shown in Fig \ref{initial}. The new domain extends
to -1 and +1 (-50\,pc to +50\,pc) in all three Cartesian directions. The pressure
and hence density outside the extracted region is defined
by matching the average pressure across the central region to the equivalent 
stable equilibrium state density of n$_H$ = 0.643 cm$^{-3}$ with the same 
pressure. The pressure-equivalent stable and unstable states are explicitly indicated in Fig \ref{eqm}. 
The surrounding medium is
uniform and stationary. The total mass in the entire domain is approximately
20,000\,M$_\odot$, of which 2800\,M$_\odot$ is in the extracted region, $r\leq0.5$.
Note from Fig \ref{initial} that the narrow density range results in equally narrow
ranges of pressure and temperature. Extracting a spherical volume from a larger
simulation in this way results in the edge of the volume cutting through some
structure, but the range of density and pressure is low enough that this has
little effect. The gravitational potential is also smoothly 
symmetric and relatively shallow. It is against this potential background of the
developing cloud complex that deep localised wells should form, if the 
subsequent clumps in the complex undergo gravitational
collapse. 

It is also useful to consider the other stable density matching this
pressure, specifically n$_H \approx 100$ cm$^{-3}$, as indicated in Fig \ref{eqm}. 
Assuming isobaric evolution, 
this is the highest density that can be achieved
by thermal instability alone with this choice of heating and cooling techniques and
parameters. In reality, 
the pressure which is roughly the same across the cloud, reduces with
time, so this provides a first order
estimate in monitoring when gravity begins to become important: densities above 
this level indicates gravity taking over from the thermal instability.

\section{Numerical methods and model}\label{methods}

The MG HD code has been used here in the same manner as throughout 
Papers I to IV. It is an upwind, conservative shock-capturing scheme employing multiple 
processors through parallelisation with the message passing interface (MPI) library. 
Integration in time proceeds according to \cite{falle91} using a second-order accurate 
Godunov method \citep{godunov59} with a Kurganov Tadmor \citep{kurg00} Riemann solver.
Self-gravity is computed using a full-approximation multigrid to solve the Poisson equation.
The same hierarchical AMR method \citep{falle05,hubber13} is employed on an 
unstructured grid. Free-flow boundary conditions were imposed 
on all boundaries. For more details, please consult Papers I to IV and Paper I in particular.

Table \ref{table-sims} presents details of the simulations presented
in this work, as well as the Paper IV HD simulation.
The HD simulation in Paper IV employed 8 levels of AMR, with 
$8\times8\times8$ cells on the coarsest G0 grid over a domain 300\,pc
on a side, making the finest grid resolution available 0.29\,pc on G7. G0 needed to
be coarse to ensure fast convergence of the MG Poisson solver. The
central region of this simulation was extracted in order to provide the initial condition
for resimulation, in Models 1 to 4 in this paper. The Model 1 simulation 
herein defined $10\times10\times10$ cells on the coarsest G0 level and 5 further levels of AMR.
Mapping of the region extracted from the Paper IV HD simulation onto the Model 1 simulation 
was performed in a simple linear fashion over all three coordinate directions
for every cell in question. The diffuse structure in this grid was well-resolved (by 10 or 
more cells) in order to ensure no loss of detail, as can be seen in Fig. \ref{initial}. 
As can be seen from Table \ref{table-sims}, the Model 1
simulation has approximately the same finest physical resolution as the
original HD simulation from Paper IV. As the Model 1 simulation
evolved, sharper, higher density features formed. To better resolve
these features new simulations with an additional 1 or 2 levels of AMR
were started from various times (Models 2 and 3, respectively). The
Model 3 simulation would have $1280^3$ cells on its finest level if it
were fully resolved. A higher resolution Model 4 simulation added a 9th level of AMR to resolve structure at 
0.039\,pc, but this was only briefly used due to the high computational cost. The simulations
were either increased in resolution from an earlier stage, or stopped, when the maximum
densities no longer met the Truelove criterion \citep{truelove97}. Finally, to examine
the final evolution of the most massive clump, an extra Clump resimulation extracted the $0.2^3$ region
($10 \times 10 \times 10$\,pc) containing this clump and resimulated it at a resolution of 0.016\,pc in order
to investigate gravitational collapse of this individual clump, details
of which are provided on the final line of Table \ref{table-sims}. The
simulations were typically run across 96 cores of the ARC2-MHD DiRAC1 HPC facility at the University
of Leeds (see also the Acknowledgments) and used in excess of 100,000 CPUhours. 

\subsection{Heating and cooling prescriptions}

The same heating and cooling prescriptions from Papers I to IV were employed, which
are now summarised in brief.
A constant heating factor of $\Gamma = 2\times10^{-26}$ erg\,s$^{-1}$ was used, with the
heating rate equal to $\Gamma \rho$. For low-temperature cooling ($\leq10^4$ K), the detailed 
prescription of \cite{koyama00}, fitted by \cite{koyama02}, and corrected according to 
\cite{vazquez07} was used. At temperatures above 10$^{4}$\,K the CLOUDY 10.00 prescription 
of \cite{gnat12} was used. These choices have enabled the definition
of cooling rates over the temperature range from 10\,K to 10$^8$\,K - the range required 
by subsequent feedback simulations. The complete cooling prescription has been efficiently implemented 
as a lookup table. The neglected processes and simplifications in this approach are discussed
at length in Paper I.  

It is worth emphasizing here that a different choice of
heating and cooling prescriptions can give very different results, in some cases
suppressing the unstable region almost entirely 
\citep[see][and their Fig 4 in particular]{wolfire95}. \cite{micic13} studied
the influence of the choice of cooling function on the formation of molecular clouds in high-resolution 
three-dimensional simulations of converging flows. They found that a number of the cloud properties, 
such as the mass and volume filling fractions of cold gas, are relatively insensitive to the choice of 
cooling function. On the other hand, the cloud morphology and the large-scale velocity distribution 
of the gas do strongly depend on the cooling function. They also investigated the properties of the 
dense clumps formed within their cloud. They found that the majority of these clumps are not 
self-gravitating, suggesting that some form of large-scale collapse of the cloud may be required 
in order to produce gravitationally unstable clumps and hence stars. It remains to be explored
in future work the extent to which different heating and cooling prescriptions affect this work.
The prescriptions used herein, that study precisely
such a large-scale collapse as referred to by Micic et al., 
are appropriate for the column density of the initial diffuse atomic cloud in our simulations.

\section{Resulting evolution of the cloud}\label{evolve}

The evolution of the cloud and the properties of the structure formed 
across the cloud complex are discussed in this Section. In particular, time-variation of 
properties and derived
statistics as well as time-snapshots of slices, projections and isosurface plots reveal
these properties. Plotting mechanisms within the MG code are used as well as the 
visualisation software VisIt \citep{visit}.

\subsection{The dynamic evolution of the cloud}

\begin{figure*}
\centering
\includegraphics[width=\textwidth]{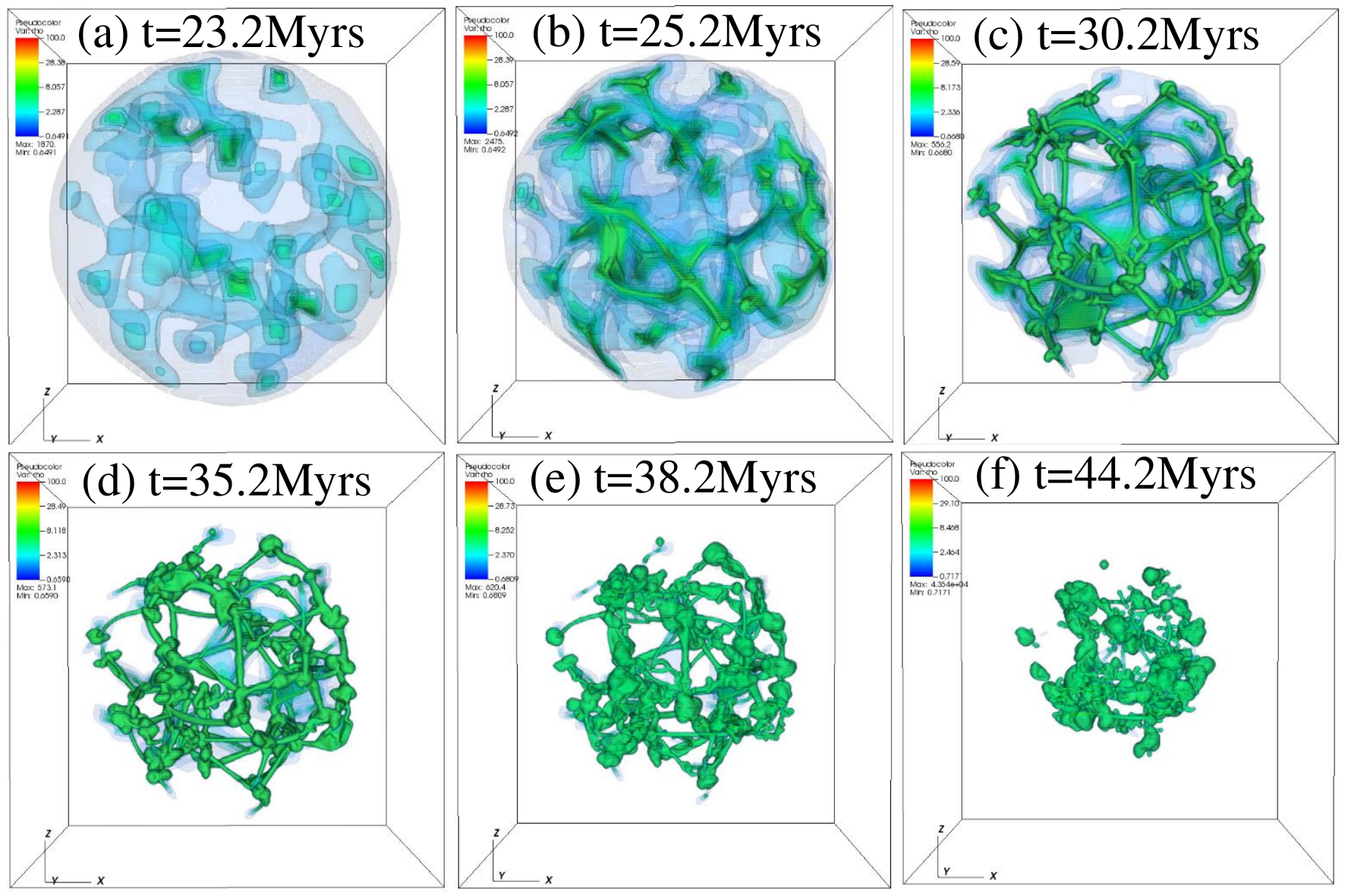}
\caption{Evolution of the diffuse cloud into a complex of clumps
shrinking through gravitational collapse. Six time snapshots from
Model 3 are shown, in a box with sides of length 50\,pc. Each snapshot shows a 3D isosurface 
rendering of density, with ten isosurfaces shown at density levels
ranging linearly up to n$_{\rm H}$=100\,cm$^{-3}$. Isosurface 
opacity increases with density, reaching high opacity by 
n$_{\rm H}$=10\,cm$^{-3}$, sharper and
coloured green (digital-only) in the figure. Lower
density more translucent isosurfaces are diffuse and bluer in colour (digital-only), indicating
the extent of unstable material ($1<$n$_{\rm H}<$10\,cm$^{-3}$).
The isosurfaces have been created from Silo datafiles using the
VisIt toolkit. Movie and Silo format raw data files are available 
from https://doi.org/10.5518/483 .}
\label{evolution}
\end{figure*}

Fig. \ref{evolution} provides an overview of the evolution of the cloud, from when the 
first condensations cause steep density increases at t=23\,Myrs to the gravitational 
contraction of the cloud complex as a whole, on a time-scale of the 50\,Myr free-fall time.
For comparison, the timescale of the maximum growth rate of
the thermally unstable initial condition with zero thermal conduction is 5.6\,Myrs 
\cite[as calculated using eq. 31 of][]{field65}.

Strikingly evident from this figure is the formation of obvious filaments, existing with high 
densities over 100\,cm$^{-3}$ for more than 10\,Myrs. These filaments connect the clumps that 
are condensing from the unstable medium at the highest density locations. Over time, the 
clumps grow in size and mass and as the volume of the cloud complex contracts under
gravity, effectively absorb the filamentary network that connected them. This is precisely
the accepted picture of star formation outlined in the introduction to this paper. 
Here, this state has evolved naturally from a diffuse stationary initial 
condition, under only the influence of thermal instability and self-gravity. 

Fig. \ref{evolution} also reveals the long-lasting 
nature of the filaments - once they have formed, around 25\,Myrs into the simulation, they 
exist for the next 15\,Myrs. Measured widths
increase over this period as the filaments accrete material, although lengths
reduce as the clumps approach one another during the gravitational collapse of the whole cloud.
There is no evidence in this hydrodynamic simulation of striation-like structure perpendicular
to the filaments. This is in agreement with the fact that striations are thought to arise from 
the effect of magnetic fields \citep{tritsis16}. We see such magnetically-aligned striations 
in our MHD simulations (e.g. as can be seen in fig. 3 of Paper IV) and will explore their
origins in the follow-on equivalent magnetic paper (in prep.) to this hydrodynamic paper.

\begin{figure*}
\centering
\includegraphics[width=0.97\textwidth]{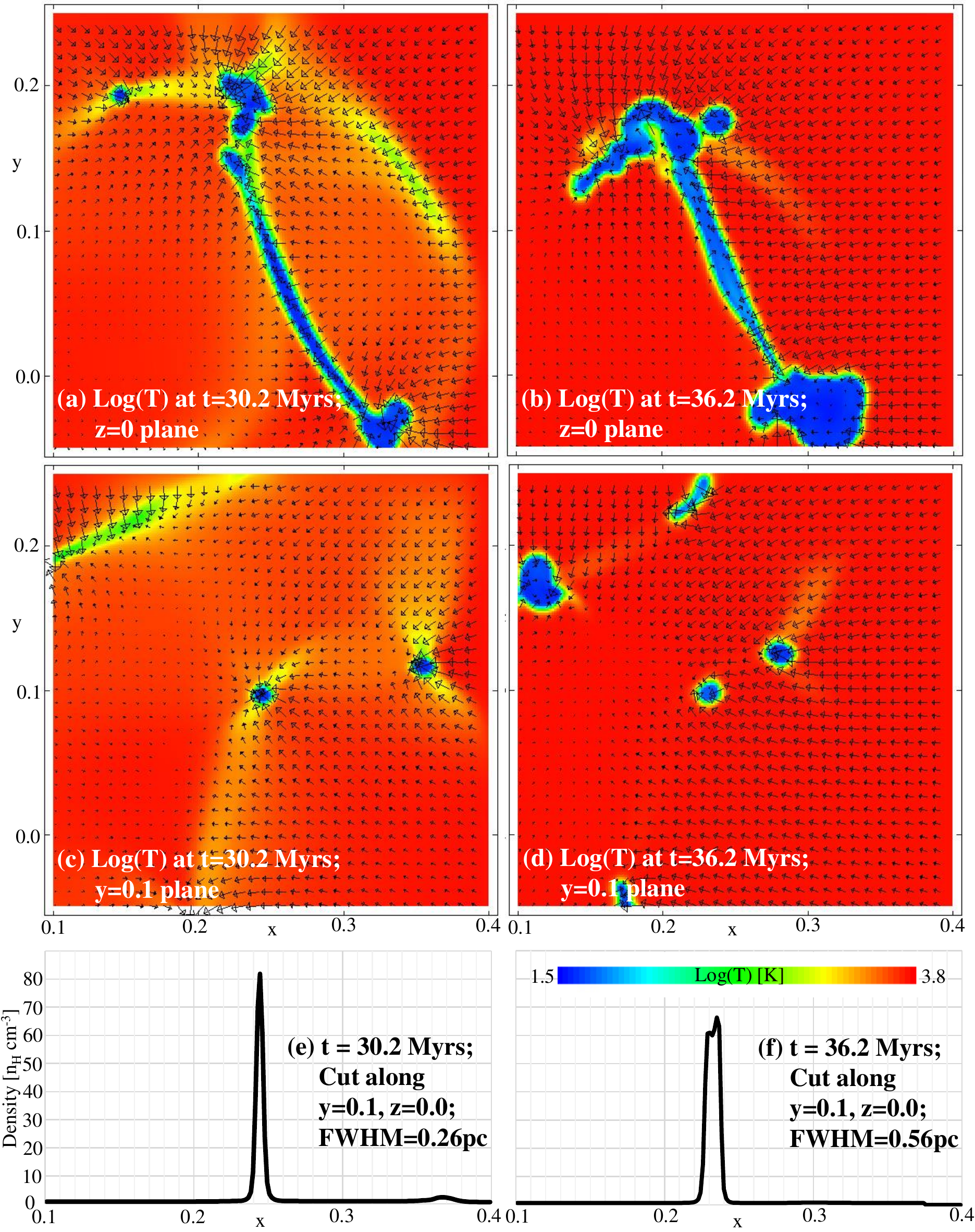}
\caption{Detail of a particular filament and its appearance at 30.2\,Myrs and 36.2\,Myrs.
(a) and (b) show slices of temperature on the $z=0$ plane revealing flow onto and along
the filament at 30.2 and 36.2\,Myrs respectively. (c) and (d) show slices of temperature
perpendicular to this on the $y=0.1$ plane at the same times revealing flow towards the 
filament and its short extent in the $x$ and $z$ directions compared to the elongation 
shown in (a) and (b). (e) and (f) show cuts in density along $y=0.1$, $z=0$ revealing
the sharply peaked nature of the filament and its width. Raw data is available from 
https://doi.org/10.5518/483 .}
\label{filament}
\end{figure*}

The filamentary nature of the cloud in this hydrodynamic simulation
came as somewhat of a surprise. Our previous work \citep{wareing16} had revealed filamentary structure in 
projection of MHD simulations, but no clear indications of filaments in any of the purely HD 
simulations. That previous work had a base level resolution of 0.29\,pc, at least four times 
lower than that shown in Fig. \ref{evolution} and comparable to the largest filamentary widths 
measured from Model 3. Any filaments in earlier HD work should have been at the very limit of detection.
Even so, reexamination of the hydrodynamic simulations presented in Paper I 
reveals that early in their evolution, the numerous clumps forming across the cloud have
a tendency to be elongated toward each other. Isosurface rendering has shown this nature
more clearly than column density projections shown in Paper I, which are uniformly clumpy.
The existence of elongated structure, even at the grid-scale in these early simulations, lends
credibility to their convincing detection at the higher resolutions employed here.

It is now prudent to investigate these filaments further. In Figure \ref{filament} we show
slices and cuts through the domain that illustrate the structure and flow in a long-lived filament.
It is clear from comparing planes at $z=0$ to perpendicular planes at $y=0.1$ that
these structures are indeed extended linear filaments, several pc long compared to their 
sub-pc widths. The spacing between the high density regions is 5-10\,pc, which is of the 
order of the acoustic  length scale, defined as the sound speed multiplied by the cooling 
time. These regions then evolve into the filaments and clumps as the thermal instability 
develops. Unstable gas flows onto the filament and over
time they grow in width, from 0.26\,pc at 30.2\,Myrs to 0.56\,pc at 36.2\,Myrs in the 
filament shown here (FWHM measurements of the density peaks in (e) and (f)). The radius 
is not a property of the thermal instability, but instead is defined by the amount of unstable 
gas that can fall in due to the flow induced by the thermal instability. All the same, these widths are in 
reasonable agreement with the observed range of filamentary widths \citep{pano17} and the
debated characteristic width of filaments \citep{arzou18}, especially bearing in mind the 
difference between the density peaks measured here and observational measurements
based on molecular emission. It is worth repeating here that 
the resolution of the simulation presented in Fig. \ref{evolution} is 0.079\,pc. The Model
4 simulation goes down to 0.039\,pc in high resolution studies of this period of the cloud 
evolution. The grid resolution is considerably less than the measured width of the filaments 
in the simulation - they are not forming on the grid scale of these simulations.

Velocity vectors on the planes in (a) to (d) illustrate the nature of the flow. 
The flow onto the filaments from the surrounding medium is clear, with velocities
on the order of a few km\,s$^{-1}$, but careful
inspection of the velocity vectors inside the filament shown in (a) and (b) reveals flow
along the filaments at magnitudes of less than 1 km\,s$^{-1}$ towards the growing 
clumps at either end. This is in good agreement with observations \citep{peretto13}.
Further work is now planned to explore the mass-flux onto and along these filaments 
and the difference introduced by magnetic fields of various strengths.

During the early evolution of the cloud, the over-dense regions present in Fig. \ref{initial}
increase in density, and then briefly stretch into sheets before becoming filaments, whilst a clump grows at their 
centre. Examination of the pressure distribution in Fig. \ref{initial} illuminates some
reasons for this evolution. In the unstable range, regions of 
higher density have lower pressure, hence the flow  of material from high to low  pressure increases 
the density further and pushes the region into cold stable phase. Reducing density at the high pressure
locations, as a consequence of material flowing away from these regions, stabilises tenuous 
low-density warm material. This is analogous to the formation of a foam under phase transition, 
where the condensed liquid exists in an interconnected filamentary network with bubbles of gas - 
bench-top X-ray tomography of a monodispersed liquid foam has revealed a strikingly similar 
structure \citep{meagher11} to that which is produced here and shown in Fig. \ref{evolution}.
There are also similarities to Voronoi foam, which has been applied at 
cosmological scales to models of galaxy distributions \cite[see, for example,][and references to this work]{icke91}.

\begin{figure}
\centering
\includegraphics[width=85mm]{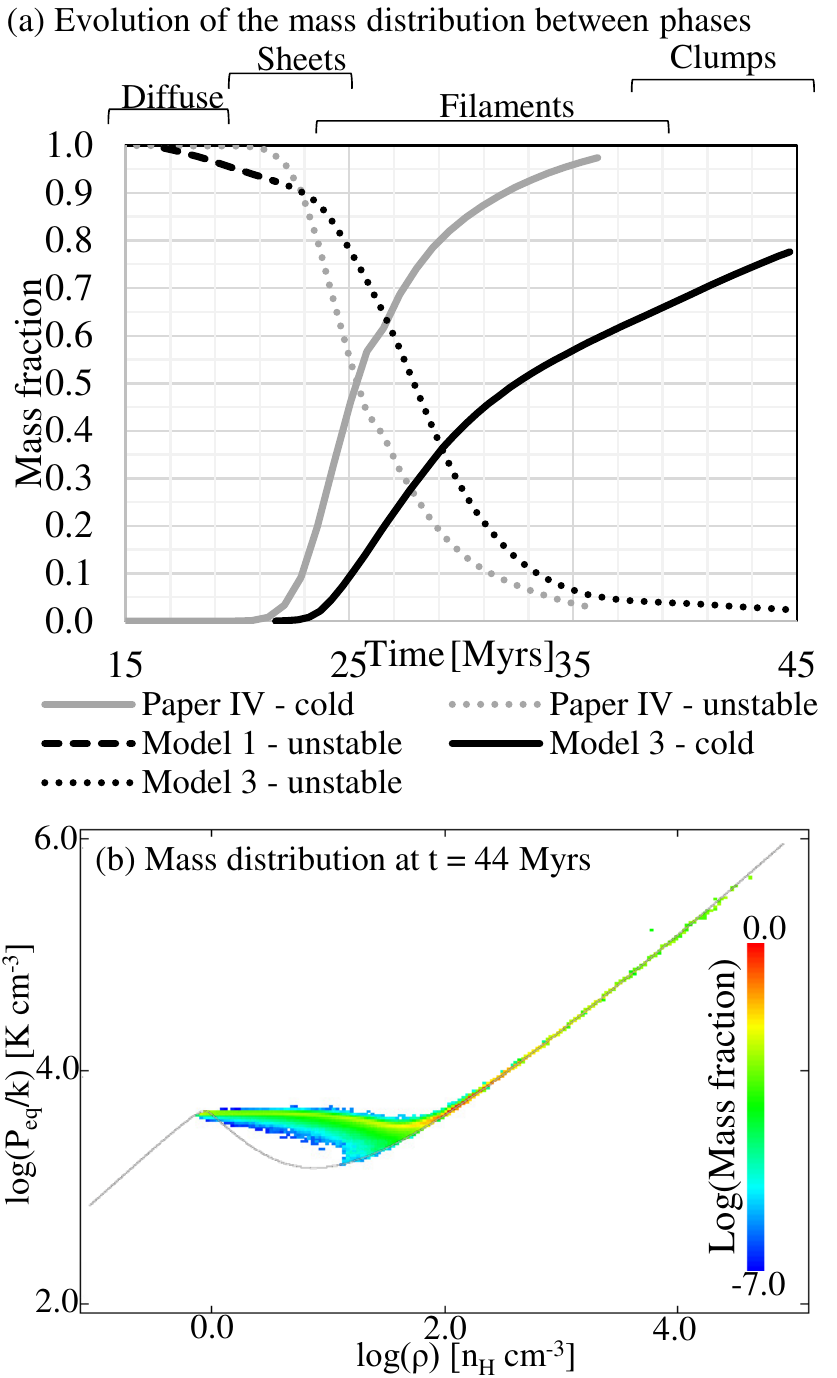}
\caption{(a) Evolution of the mass 
in the appropriate temperature range (cold\,$<$\,160\,K, 160\,K$<$\,unstable\,$<$5000\,K, 
warm\,$>$5000\,K), with indications of the dominant structure at that time above
the line graph. (b) Mass
distribution of the cloud material in the Model 3 simulation at t=44.2\,Myrs, as shown
in Figure \ref{evolution}(f). The grey line indicates the equilibrium curve for our
choice of heating and cooling prescriptions. Raw data is available from 
https://doi.org/10.5518/483 .}
\label{phases}
\end{figure}

\subsection{Evolving properties of the cloud}

Fig. \ref{phases}(a) gives mass in the appropriate temperature 
range, which does not mean that it is on the equilibrium curve, for a subset of the time evolution. 
The first 15\,Myrs of the evolution are not shown as nothing varies. The filamentary 
network that develops is characteristic of the transitionary phase as the simulation
evolves the unstable diffuse material into stable states. 
As the majority of the mass stabilises into either warm or cold states,
the filaments are absorbed
into the individual cold clumps of the collapsing cloud complex. These individual clumps
have much higher densities and hence have the potential to collapse under gravity and 
form stars on shorter timescales than the whole cloud complex.

The mass distribution of cloud material at t=44\,Myrs in the
Model 3 simulation is shown in Fig. \ref{phases}(b). It is clear from this plot that material still
persists out of equilibrium in the unstable density range. Analysis of the mass
distribution of material by density range reveals that approximately 10\% of the cloud material 
is in the density range from 1 to 10\,cm$^{-3}$ (c.f. for the Paper IV HD simulation, the fraction
is approximately 35\% at t=36\,Myrs - considerably more than the fraction of 5\% implied from
Fig. \ref{phases}(a) at this time). As this material is out of equilibrium, it
is difficult to classify and nominally should be assigned to its final state, as we have
done here using ranges in temperature. Hence, the mass fraction of unstable material in equilibrium is low, as shown in Fig. 
\ref{phases}(a), but the volume-filling-fraction of this low-density non-equilibrium material
is comparatively high. Observations have confirmed that such thermally unstable material can be
detected in the temperature range 400 to 900\,K \citep{begum10}. Whilst the model herein
is clearly able to dynamically produce thermally unstable gas, further work
is required to compare this in detail to observations. Fig. \ref{phases}(b)
also shows the high density material at high pressure in thermal equilibrium, now collapsing 
under gravity in the individual cold dense clumps of the complex.

\newpage

\begin{figure}
\centering
\includegraphics[width=80mm]{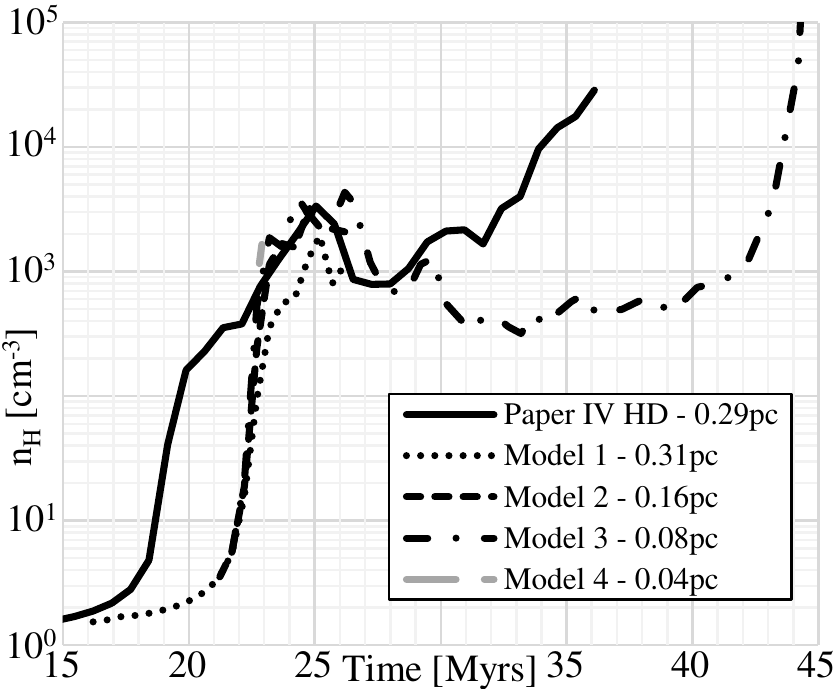}
\caption{Evolution of the maximum density in the domain. The faster
  formation of dense clumps in the Paper IV HD simulations is due to
  the compression at the outer surface of the cloud, which results
  from our choice of initial conditions (see main text).
Raw data is available from https://doi.org/10.5518/483 .}
\label{maxrho}
\end{figure}

The variation of maximum density with time is shown in Fig \ref{maxrho}, for the same subset
of the time evolution. The time indicated on the horizontal axis is that elapsed since the
start of the Paper IV HD simulation; the first 15\,Myrs are similarly not shown as the maximum 
density is close to the initial density in the diffuse atomic cloud.
The black solid line indicates the trend of maximum density from the Paper IV HD simulation.
The steep rise from 18 to 21\,Myrs is due to condensations
close to the edge of the cloud that have been influenced by the pressure difference
caused by the reducing average pressure across the cloud compared to that of the surrounding 
medium. Whilst these may or may not be representative of reality, in this work the
study of thermal instability and gravity is under examination, hence the isolation
of the central region of the Paper IV HD cloud extracted and used
to define the initial condition of the Model 1 simulations. This is also the reason for the 
{\it reduction} in the maximum density present in the domain at the initialisation of the 
Model 1 simulation and the steep rise in density shown after 30\,Myrs 
by the Paper IV HD simulation, which is the gravitational collapse of the condensations
close to the edge of the cloud.

The maximum densities in the Model 1-4 simulations are
representative of structure solely generated from the thermal instability. These
simulations show that thermal instability triggers a steep rise in density
around 22\,Myrs, as the filamentary network forms. During this steep rise, at 21.7\,Myrs
the Model 2 and Model 3 simulations are both  
started by adding extra levels of AMR in order to ensure resolution of later
structure; the starting of simulations with extra levels of AMR is earlier than the maximum density peaks
as adding extra levels of AMR once already
at peak density risks loss of unresolved structure in the run up to that time. 

By 23\,Myrs, the
maximum density in the Paper IV HD simulation is back in agreement with the 
maximum densities present in the Model 1, 2 and 3 simulations.
The maximum density in the Paper IV simulation is now representative of thermal instability-driven
structure in the extracted central r=0.5 (25\,pc)
region of the cloud used to initialise the Model 1 simulation, rather 
than condensations close to r=1.0 that were unduly influenced by 
the pressure difference between cloud and surroundings. 
These condensations lead to the marked difference between the Paper 
IV HD simulation and all the other simulations in Fig. \ref{maxrho}. 

By 25\,Myrs,
the maximum density in all simulations is in agreement and reaches n$_{\rm H} \approx 3000$ cm$^{-3}$. A detailed 
inspection of this behaviour (through adding an extra level of AMR at t=22.8\,Myrs to
initialise the Model 4 simulation) indicated that it {\it does not} lead to gravitationally
bound clumps that may subsequently form stars. Inspection of the velocity field
around these structures reveals velocities on the order of a few km\,s$^{-1}$ that 
are non-convergent, resulting from flow of thermally-stable material from warm, tenuous 
to cold, dense conditions. Kinetic energy dominates and as a result,
these structures break apart on relatively short timescales compared to their free-fall times.
The long duration of the relatively high maximum density in Fig \ref{maxrho} 
from 24 to 27\,Myrs is indicative of the period over which the filamentary 
structures form and evolve in the cloud. It is not the peak density of individual clumps
that may collapse under the effect of gravity. 
The Model 3 simulation resolves the peak in maximum densities into several separate 
peaks for this reason.

The importance of correctly identifying this period of dynamically-dominated structure 
in any simulations cannot be over-emphasized. Dynamical and non-equilibrium
effects have been shown to be important for cloud evolution \citep{goi16}. 
Mechanisms for forming star particles could mis-identify these
regions of high density, as they meet conditions for inward flow,
potential minimum and Jeans instability. Depending on the frame of reference
selected for calculating  energies, the balance between kinetic,
gravitational and thermal energy can also be misleading, an issue that has also
been noted in observations \citep{ball18}. Followed 
carefully at high resolution, they do not collapse under their own self-gravity. 

\begin{figure*}
\centering
\includegraphics[width=170mm]{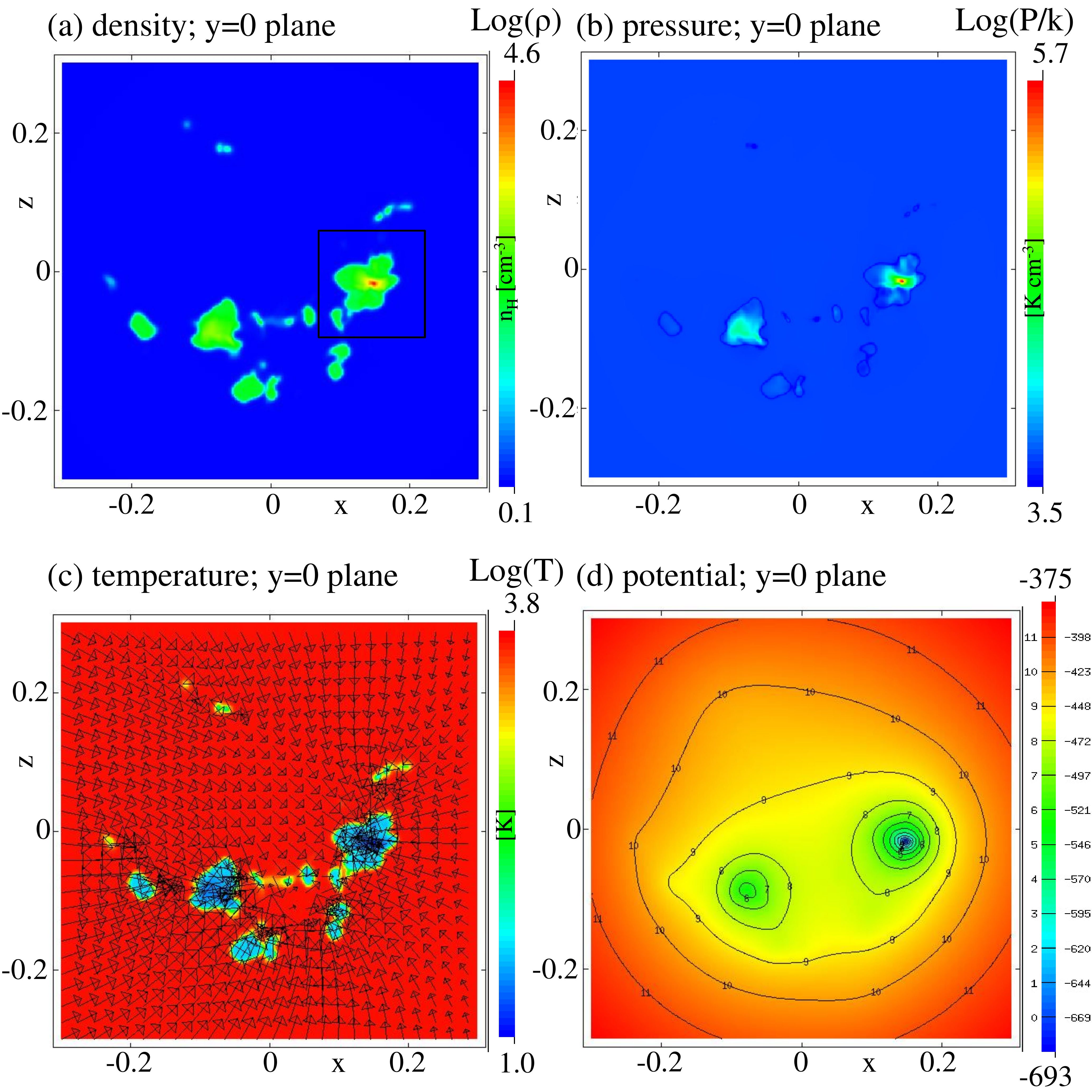}
\caption{The evolved simulation, showing a slice through
the domain at y=0.11328 through the densest clump in the grid.
Shown are (a) density, (b) pressure, (c) temperature and (d)
gravitational potential of the Model 3 simulation at t=44\,Myr. 
Vectors of velocity, scaled on the 
largest magnitude in the slice of 2.5\,km\,s$^{-1}$, are shown
in panel (c). The box in panel (a) indicates the volume, centred
on the most massive clump, resimulated later at higher resolution.
The unit of distance is 50\,pc.
Raw data is available from https://doi.org/10.5518/483 .}
\label{wholecloud}
\end{figure*}

The original Paper IV HD simulation contained sharp unresolved structures at this time
and did not resolve multiple dynamic peaks around 25\,Myrs, but did
capture the transience of this period, reducing back to lower densities in the cloud
until the condensations at the edge of the cloud begin to collapse around 29\,Myrs. 
The resolved Model 3 simulation captures the true collapse of quiescent thermal instability-driven structure
beyond this point showing that longer-lasting filamentary structure forms with densities on the
order of several hundred per cubic centimetre. By 40\,Myrs or so
individual clumps have gathered enough mass to begin to collapse under the influence
of gravity. This is against the background of the infall towards the deep central
potential well at the centre of the domain.
There is therefore a long period of time (up to 10\,Myrs from t=28\,Myrs) where the cloud 
consists of quiescent high density clumps connected by lower density filaments that 
have not collapsed under gravity. These clumps
have size scales on the order of a few pc and are therefore not diffuse microstructure
in the ISM \citep{stanimirovic18}, but do resemble pressure confined clumps
noted as a pre-requisite for star formation \citep{kainu11}.
Whilst we do not track the formation of molecules, the timescale here is not
dissimilar to the $\sim$10\,Myr time scale of cloud formation recently derived for formation of
molecular hydrogen in an isolated dark cloud \citep{zuo18}.

\section{An investigation of the cloud at t=44\,Myrs}\label{snapshot}

\begin{figure*}
\centering
\includegraphics[width=\textwidth]{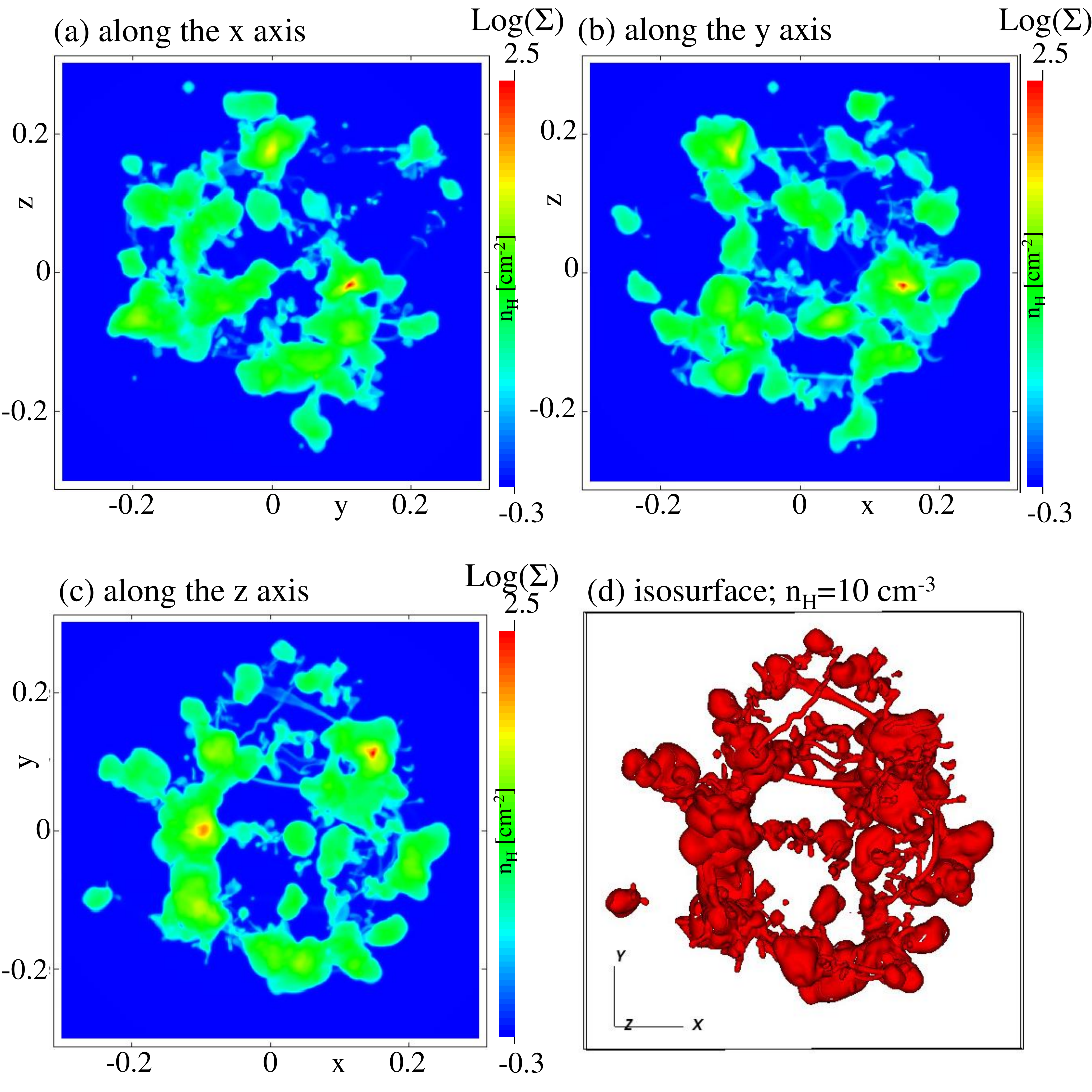}
\caption{Projections of column density in the three axial directions and an isosurface
of density, rotated to match the projection along the $z$ axis for side-by-side
comparison, of the Model 3 simulation at t=44\,Myrs as in the previous figures.
The unit of distance is 50\,pc.
Raw data is available from https://doi.org/10.5518/483 .}
\label{coldens}
\end{figure*}

The appearance of the cloud as a whole at t=44\,Myrs is shown in Figs. \ref{wholecloud}
and \ref{coldens}. This is the time at which 
the majority of the material in equilibrium in the domain has stabilised 
and can be considered the thermal end-state of the cloud. The remaining unstable material
is in thin layers surrounding the clumps. Few filaments remain. In the unrealistic scenario that
stars do not form, the future of the cloud is collapse under gravity on the free-fall timescale
of the cloud. The question is whether individual clumps can or are collapsing on shorter 
time scales. Slices of density, pressure, temperature and potential through the 
simulation domain show the clumpy nature of the cloud complex, as well as the fact 
that only the most massive clumps have increased central pressures at this time. The 
temperature slice shows both temperature on the colour scale and vectors of total 
velocity projected onto this plane. The uniformity of the warm stable surroundings is 
immediately apparent - there is effectively very little variation across this intra-clump 
medium, even though the flow field is complex. The velocity vectors show there 
are inflows towards the individual clumps apparent on this plane, as well as the overall
infall towards the centre of the most massive clump. The depth of the potential well, as 
well as the previously noted smooth nature of the distribution is apparent from the slice 
through potential. These slices through the domain do not show any filamentary
structures. At this time, they are now rare, having been absorbed into clumps, or
dissipated along their length into chains of smaller clumps. Fig. \ref{coldens} shows
projections of column density in the three axial directions and a density isosurface of 
the Model 3 simulation, revealing the few remaining filaments. In particular, in the upper
right of panel (a), the break-up of a filament into a series of clumps along its length
appears to be occurring at this late time.

\subsection{Analysis of the clumps across the cloud}

\begin{table*}
\begin{center}
  \caption{Properties of the 21 clumps with more than 20\,M$_{\odot}$ identified by 
  the FellWalker algorithm, at t=44\,Myr in the Model~3 simulation. Snapshots of slices through the clumps are available
  from  https://doi.org/10.5518/483 .}
  \label{table1}
  \begin{tabular*}{\textwidth}{lllllllllll}
\hline
 &  M$_{total}$ & M$_{warm}$ & M$_{unstable}$ & M$_{cold}$ &  $\rho_{max}$ & T$_{min}$ & Scale & v$_{max}$ & v$_{min}$ & Notes\\
 & [M$_{\odot}$] & [M$_{\odot}$] & [M$_{\odot}$] & [M$_{\odot}$] & n$_{\rm H}$ [cm$^{-3}$]& [K] & [pc] & [km\,s$^{-1}$] & [km\,s$^{-1}$]\\
\hline
 A & 2.37e2  & 5.39e1 & 3.19e0 & 1.78e2 & 4.62e2 & 30.5 & 3.5 & 2.2 & 0.11 & Spheroidal, extended arms\\
 B & 2.64e2  & 2.36e1 & 4.10e0 & 2.37e2 & 9.92e2 & 21.5 & 4.0 & 2.8 & 0.31 & Prolate spheroid \\
 C & 2.71e2  & 4.49e1 & 4.25e0 & 2.22e2 & 7.76e2 & 22.9 & 4.0 & 2.4 & 0.12 & Pyramidal\\
 D & 3.54e2  & 5.65e1 & 6.49e0 & 2.92e2 & 1.46e4 & 10.6 & 4.0 & 2.5 & 0.17 & Multiple lobes\\
 E & 7.27e1  & 1.05e1 & 1.01e0 & 6.12e1 & 4.16e2 & 31.3 & 2.5 & 1.9 & 0.03 & Spheroidal\\
 F & 1.08e2  & 1.55e1 & 1.01e0 & 9.12e1 & 4.32e2 & 28.8 & 4.0 & 2.6 & 0.06 & Double sphere merger\\
 G & 1.77e2  & 1.94e1 & 2.65e0 & 1.55e2 & 3.31e2 & 32.7 & 4.0 & 2.1 & 0.04 & Peanut \\
 H & 2.65e1  & 2.01e0 & 3.19e-1 & 2.42e1 & 1.54e2 & 32.4 & 2.4 & 2.5 & 0.03 & Clump on a filament?\\
I & 7.57e1  & 5.88e0 & 1.06e0 & 6.88e1 & 1.94e2 & 33.4 & 3.0 & 2.2 & 0.18 & Results of a merger?\\
J & 3.13e1  & 2.59e0 & 2.93e-1 & 2.84e1 & 2.16e2 & 32.0 & 3.0 & 3.2 & 0.08 & Co-flowing clumps\\
K & 1.04e2  & 5.10e0 & 1.14e0 & 9.76e1 & 4.18e2 & 26.9 & 2.5 & 1.8 & 0.03 & Isolated. Spherical \\
L & 2.37e1  & 1.02e0 & 3.28e-1 & 2.23e1 & 2.83e2 & 31.1 & 1.0 & 2.5 & 0.01 & Tadpole, 2.5\,pc elongated tail\\
M & 6.83e1  & 1.16e1 & 1.07e0 & 5.57e1 & 4.95e2 & 26.1 & 3.0 & 2.7 & 0.03 & Elongated. Chain?\\
N & 9.14e1  & 8.50e0 & 1.42e0 & 8.15e1 & 3.33e2 & 29.3 & 3.0 & 2.4 & 0.33 & Spheroidal, linked filament?\\
O & 4.85e1  & 8.63e0 & 1.28e0 & 3.87e1 & 2.34e2 & 29.9 & 3.0 & 1.7 & 0.02 & Spheroidal - 2 filaments\\
P & 6.84e1  & 2.20e1 & 1.66e0 & 4.47e1 & 1.89e2 & 33.1 & 3.0 & 1.9 & 0.07 & Large tadpole\\
Q & 6.63e1  & 5.19e0 & 9.80e-1 & 6.01e1 & 2.86e2 & 32.1 & 4.0 & 1.8 & 0.03 & Sph. Off-centre max rho\\
R & 2.96e2  & 3.19e1 & 2.76e0 & 2.62e2 & 3.25e3 & 16.9 & 5.0 & 2.4 & 0.13 & Multiple lobes - subclumps?\\
S & 7.25e1  & 7.70e0 & 1.00e0 & 6.38e1 & 3.92e2 & 27.4 & 5.0 & 2.5 & 0.02 & Double merger\\
T & 3.57e1  & 1.53e0 & 7.64e-1 & 3.34e1 & 2.06e2 & 33.0 & 3.0 & 2.3 & 0.03 & Prolate spheroid\\
U & 3.36e1  & 3.72e-1 & 3.91e-1 & 3.28e1 & 2.03e2 & 33.6 & 2.5 & 2.7 & 0.15 & Spheroidal, sub-clumps\\
\hline
  \end{tabular*}
\end{center}
\end{table*}

It is now meaningful to identify and analyse massive clumps in the simulation.
The FellWalker clump identification algorithm \citep{berry15} has been implemented
into MG in order to do this. Berry described FellWalker as a watershed algorithm that 
segments multi-dimensional data above a pre-set background level into a set of disjoint 
clumps, each containing a significant peak. It is equivalent in purpose to the CLUMPFIND
algorithm \citep{williams94}, but unlike CLUMPFIND, FellWalker is based on a 
gradient-tracing scheme which uses {\it all} the available data, `walking' from
each datapoint according to the steepest gradient to the peak associated with 
that gradient. Berry 
performed comparisons with CLUMPFIND and showed that the results produced by
FellWalker are less dependent on specific parameter settings than are those of
CLUMPFIND.

Whilst Berry designed FellWalker to identify significant peaks in density, here
FellWalker has been implemented to identify significant wells in gravitational
potential. The gravitational potential has a significantly smoother distribution 
than the mass distribution and hence makes an excellent choice for
identifying clumps directly from the simulation without any parameterised 
smoothing step. After implementation, the algorithm was thoroughly tested
with simple 2- and 3-dimensional potential distributions to establish confidence
in the performance and understand the optimum grid configuration for usage.
In order to achieve the best results, we applied FellWalker to the potential 
projected onto the finest grid level. The potential is sufficiently smooth for 
the projection from coarser grids to be smooth enough for FellWalker.

Applying FellWalker to the Model 3 simulation snapshot at t=44\,Myrs, the algorithm
detects 21 individual clumps with masses greater than 20\,M$_{\odot}$. This particular 
time is the time at which the highest density in the simulation, in the centre of the most
massive clump, reaches the limit of the resolution. It should be noted that this most 
massive clump is not at the centre of the simulation. The term `clump' is used by choice in order to achieve 
clarity from the diffuse initial cloud condition and the cloud complex that the combination
of these clumps forms. \cite{bergin07} review the properties of clumps based
on \cite{loren89} and \cite{williams94}: mass 
50-500\,M$_{\odot}$; size 0.3-3\,pc; mean density 10$^3$-10$^4$\,cm$^{-3}$, velocity
extent 0.3-3\,km\,s$^{-1}$; sound crossing time 1\,Myr; and, gas temperature 10-20\,K.
Properties of the 21 massive clumps identified in the Model 3 simulation are shown
in Table \ref{table1}. The range of mass, maximum density, size scale and velocity 
fit the description of Bergin \& Tafalla very well. In each clump, the majority ($>75\%$)
of the mass is in the cold phase. Minimum temperatures, which
occur in the inner regions of each clump where the lowest velocity also occurs,
are typically slightly higher than Bergin \& Tafalla's review; only the two most massive
clumps, D and R, are colder than 20\,K in their core. It is worth noting that maximum
density and minima of temperature and velocity are all co-located in these cold
($<$\,100\,K) inner regions of each clump, but often not at the centre of mass of
the clump, reflecting the complex evolution that has produced these clumps.

The maximum velocity is usually found at the edge of the clump, which is well-defined 
by a sharp gradient in temperature. Material there is falling
into the clump whilst rapidly cooling and deccelerating through the phase change - 
these are dynamic rather than stationary accreting objects. They are not simply
cold clumps stabilising out of quiescent unstable surroundings. The surroundings are in fact
in the warm stable phase and the only unstable region is across the sharply defined edge
of the clump, hence the ``unstable'' mass (defined as that in the unstable region of 
Fig \ref{eqm}, with densities between n$_{\rm H}$=1.0 and 10.0 cm$^{-3}$) is a small 
fraction of the total clump mass.

The appearance of the clumps is varied, as noted in Table \ref{table1}. Very few are 
isolated and spherical, reflective of the fact that they have 
absorbed filaments and smaller clumps. What Table \ref{table1} only partially alludes to, is the remnants
of filamentary interconnecting structures between clumps. This is present at t=44\,Myrs
as smaller condensations where a filament has dissipated along its length, extended wings 
stretching away from the clumps where filaments are being absorbed and the existence of 
isolated small clump-like structures. These small structures have masses on the order of a few solar 
masses or less and have shallow potential wells, if any discernible effect on the gravitational 
potential. In time, the isolated small structures may grow, or merge into existing clumps, 
as all are falling toward the centre of the cloud complex. At this time in the Model 3 simulation, 
there are no discernible sheet-like cloud collisions.
The subsonic velocities in the simulations here do not lead to shock-compressed thin structures.

\begin{figure*}
\centering
\includegraphics[width=\textwidth]{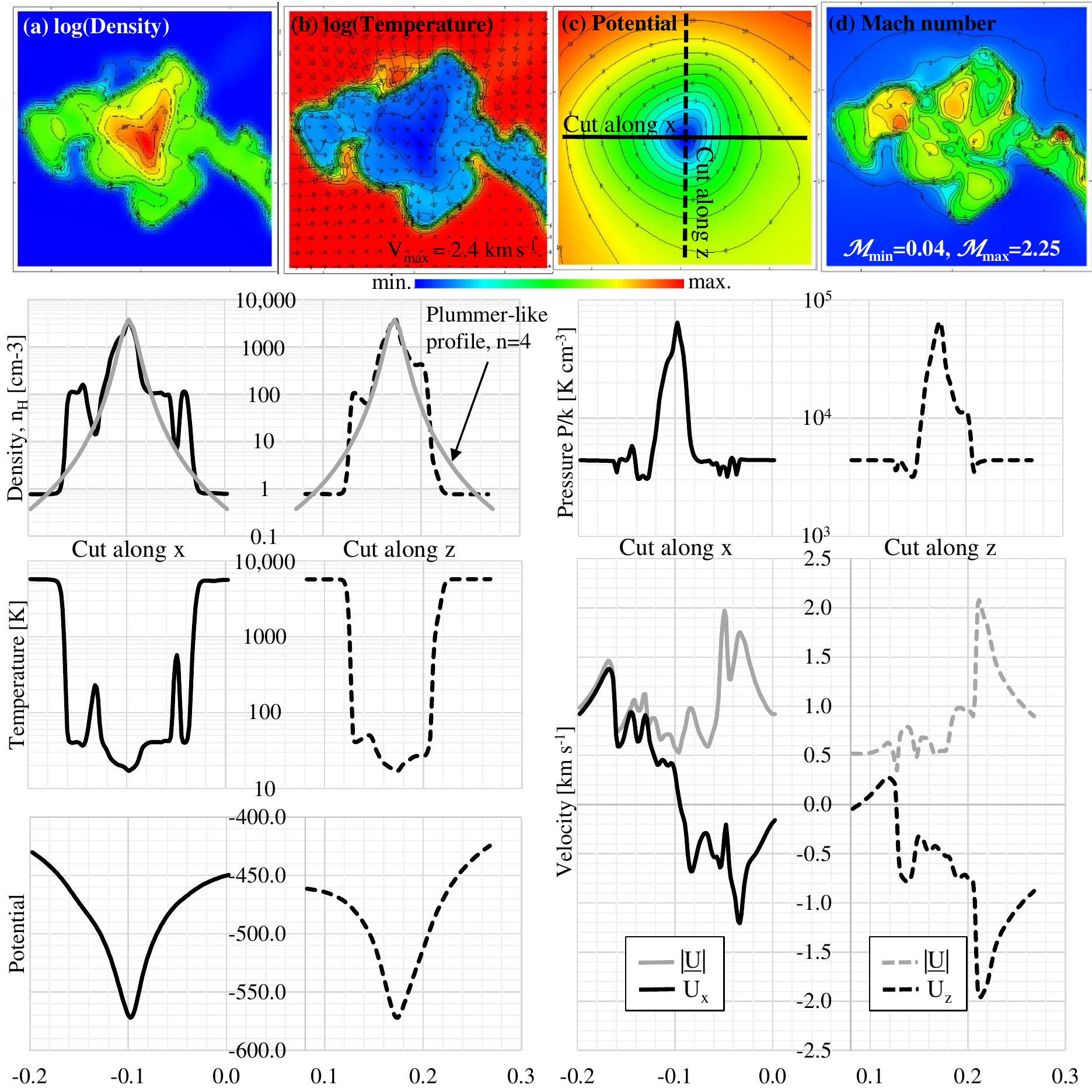}
\caption{Properties of Clump R from the Model 3 simulation at t=44\,Myr. The unit of distance is 50\,pc.
Raw data is available from https://doi.org/10.5518/483 .}
\label{slices}
\end{figure*}

The maximum and minimum velocities in Table \ref{table1} are not reflective of the
velocity dispersion of the individual clumps. The maximum velocity reflects the typical velocity
of the infalling material at the edge of the clump, uncorrected for the
velocity of the centre of the mass of the clump falling towards the centre of the cloud 
complex, up to 0.5 km\,s$^{-1}$. Further analysis has shown that the
minimum velocity is typical of the velocity dispersion of the cold material in the clump, 
on the order of 0.2-0.5\,km\,s$^{-1}$ or less. 
At the edges of the clump, the velocity of the infalling material is subsonic in the 
warm stable medium, and approaches Mach number of 1 as the material flows across the 
temperature gradient at the boundary of the clump.
This would appear to be in-line with observations of sonic velocities around clumps, which must dissipate for
the clump to eventually collapse under the effect of gravity, converting kinetic
energy into thermal and gravitational energy - the potential well deepens whilst
the pressure at the centre of the clump rises. The clumps in Table \ref{table1} 
display clear evidence of the relationship between increasing density
and deepening individual potential wells, even set against the background of the
potential of the cloud complex as a whole. Interestingly, on the smaller scale of
cores, \cite{ball18} find the extension of this from clumps to cores - the observed energy
budget of cores in recent surveys is consistent with their non-thermal motions being
driven by their self-gravity and in the process of dynamical collapse. Furthermore,
\cite{micic13} study the influence of cooling functions on the formation of
molecular clouds and find, in agreement with previous models, that the majority
of clumps are not self-gravitating, suggesting that some large-scale collapse
of the cloud may be required in order to produce gravitationally unstable
clumps and hence stars. We see that effect here and now go on to study an
individual clump in detail, followed by study of the final collapse of the most
massive clump identified in Table \ref{table1}.

\subsection{Study of an individual clump}

Turning to examine an individual clump in more detail, Figure \ref{slices} shows
slices and profile cuts through Clump R, the second most massive clump in the
Model 3 Simulation at t=44\,Myrs. The slices and profiles are cut through the
position of the deepest potential of the clump, at (-0.0977, 0.00078, 0.174).
The centre of mass of the clump is slightly offset at (-0.103, 0.00209, 0.181),
equivalent to 0.44\,pc between centre of potential and centre of
mass - further reflecting the non-symmetric, non-spherical nature of this 
clump, 5\,pc in length along its major axis. The cut along $x$ reveals multiple
peaks in density, with corresponding complex velocity structure, due to the
absorption of the filamentary network and merger with other clumps. 

It is interesting
at this point to compare the density distribution to previous assumptions
about clumps. Clearly the clump does not have a uniform density distribution.
It does, around the central peak, have a Plummer-like density distribution, indicated
in the figure by the agreement between the data (black line) and theoretical
profile (grey line). In the negative $x$ direction, this agreement stretches over
two orders of magnitude in density - on a linear $y$-axis scale, the agreement
appears very close. The comparison here is with the classic
Plummer-like profile introduced by \cite{whitworth01}, with the observationally
confined power-law index of 4, rather than the true Plummer sphere with index
of 5.  The fit takes a central density of n$_{\rm H}$ = 3780 cm$^{-3}$ from the 
data and a minimal central flat radius of 0.01 ($\approx$~0.5\,pc). The
temperature plot serves to illustrate the sharply defined edge of the clump, where
the temperature ranges quickly from the lower inner temperatures below 100\,K,
to the high temperature ($>5000$\,K) of the warm stable surroundings. Peaks
in the inner temperature profiles revealed by the $x$ cut, which correspond
to troughs in density, again show evidence for merger with earlier structure in the cloud.

The distribution of the potential, shown in panel (c) of Fig \ref{slices} is remarkable
for its smoothness compared to the complex structure apparent in density. 
In the context of simulated molecular clouds, gravitational potential is clearly
useful for identification of distinct clumps with structure-finding tools such as
FellWalker or CLUMPFIND. 

The Mach number of the flow is shown in Fig. \ref{slices}(d), with 
vectors indicating the magnitude and direction of velocity,
projected onto this plane in Fig. \ref{slices}(b). Whilst some information 
is lost in this projection, panel (b) 
makes it apparent that the highest density, coldest regions of 
the clump correspond to the lowest velocities inside the
clump. The complexity of the velocity field is apparent in Fig. \ref{slices}(d), indicating accelerating
but still sub-sonic inflow from above the clump. Note that velocity on this slice is presented
in the frame of reference of the entire simulation domain, rather than corrected
for the motion of the centre of mass of the clump towards the centre of the
slowly collapsing cloud complex. The velocity of the centre of mass
of the clump is approximately 0.5 km\,s$^{-1}$ towards the lower right corner of
the panels, indicated by the uniform velocity field in cold regions of the clump.
Clearly, velocity in the frame of the clump reduces as material
flows, under the action of the thermal instability, across the phase boundary from
warm tenuous to cold dense material. In turn, the dispersion of velocity reduces across
this boundary. Adjusting for the motion of the clump, the apparently
supersonic regions inside the clump in Fig. \ref{slices}(d) are in fact subsonic. However, the
supersonic material falling into the filament from the warm medium at the righthand edge 
of the plots is supersonic once inside the cold regions of the filament, indicative of mildly
supersonic velocities in filaments.

Of key importance in Fig. \ref{slices} are the
cuts though pressure across the clump. The central pressure is an order of
magnitude greater than the equilibrium pressure in the warm surroundings,
clearly indicating gravitational collapse. The conditions are close to thermal equilibrium,
with pressure at the centre a few percent higher than the corresponding thermal 
equilibrium pressure for that density. The central density is not in dynamical equilibrium
and is still rising as the clump continues to collapse. On the outer 
edges of the clump are rises in density and small rises in pressure, together with
variations in temperature that are indicative of the merger of clumps and absorption
of filaments. The velocities in these regions, in the frame of the clump, are sub-sonic
and so there are no shocks present inside the clump.

\newpage

\section{Power spectra of the cloud complex and an individual clump}\label{powspec}

\begin{figure}
\centering
\includegraphics[width=85mm]{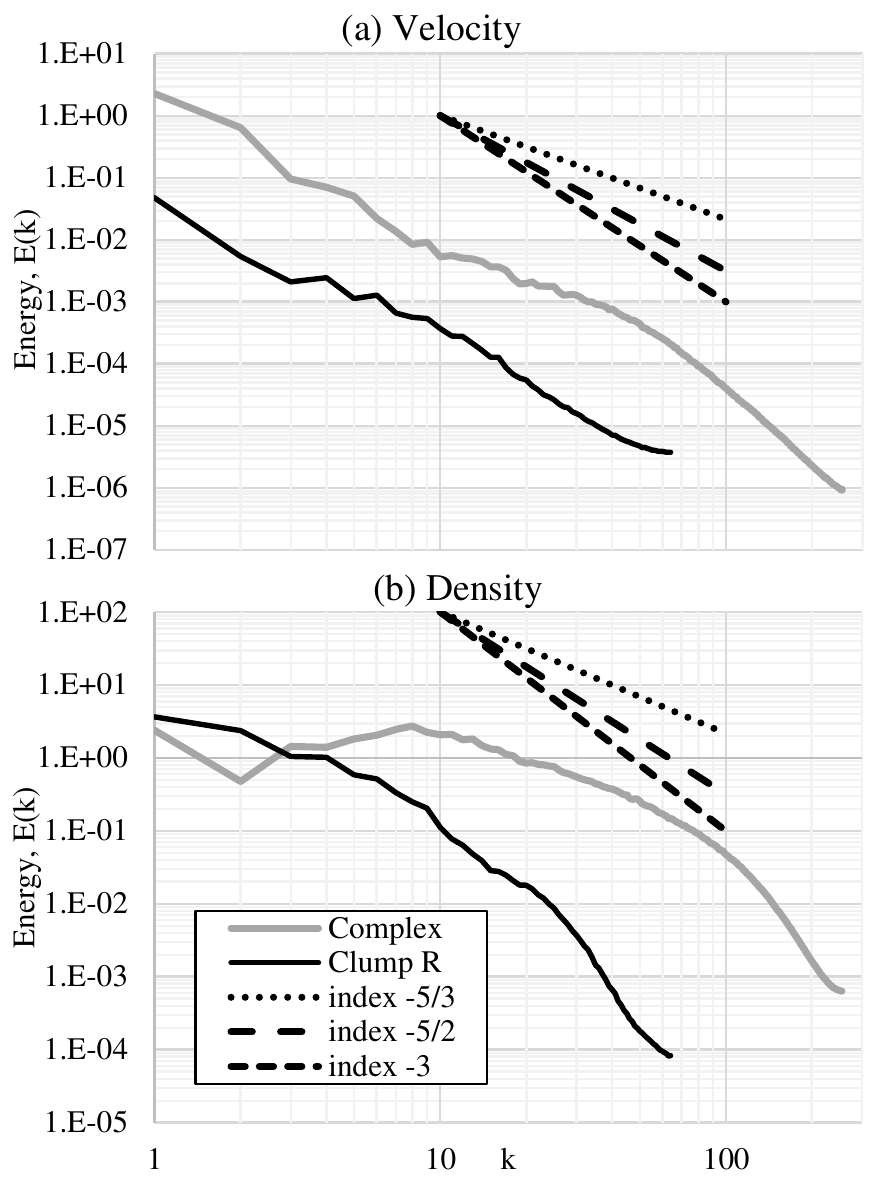}
\caption{Instantaneous power spectra of a) velocity and b) density for the cloud as
  a whole and for clump R at t=44\,Myrs. Raw data is available from https://doi.org/10.5518/483 .}
\label{spectra}
\end{figure}

In Fig. \ref{spectra} we show the snapshot power spectra of velocity and density both for the
cloud complex as a whole and for Clump R at t=44\,Myrs. The power spectra have been calculated 
from the magnitude of the complete velocity vector (v$_x$, v$_y$, v$_z$) and the density. 
There is no projection or smoothing of the velocity vector into two components on a plane
\citep[as has been shown to affect velocity power spectra by][]{medina14}. Power spectra have been calculated 
using a simple IDL\footnote{The interactive data language; https://www.harrisgeospatial.com/docs/using\_idl\_home.html}
routine which performs the Fourier Transform and bins the square of 
the 3D transform into wavenumbers to obtain the 1D power spectra. The validity of this 
method has been checked by generating density structures in 1D, 2D and 3D. These structures
have been generated using single wavenumber sine waves with varying wavenumbers and 
combinations of these sine waves with multiple component wavenumbers
to ensure the power spectra analysis returns the correct wavenumber(s) that was(were) used
to generate the density structure in the first place. 

Clear from the power spectra of velocity is that both the cloud complex and Clump R 
display an inertial range (region in wavenumber with slope of constant gradient) greater
than one order of magnitude. 
The cube encompassing the cloud complex transformed for this analysis was 40\,pc 
on a side. Between wavenumbers 3 and 30 (or physical scales from ~13\,pc down to 1\,pc), 
the spectral index is close to the Kolmogorov -5/3 spectrum generally observed in turbulence.
This is in good agreement with observationally-derived velocity power spectra, for example
an index of $-1.81 \pm 0.10$ derived by \citet{padoan06} for the Perseus molecular cloud
complex. 
At $k=30$ there is a break in the spectrum, to an index steeper (less than) than -3 
implying rapidly reducing kinetic energy with decreasing physical size scale. The validation tests
showed that single precision numerical noise generated an index of -5 when a pure, single wavenumber
sinewave was used to generate 3D input data, tending to indicate therefore
there is still some power at these scales as the spectrum is not as steep as -5.
Calculation of power spectrum at earlier times shows that before 23\,Myrs, the 
index of the spectrum is steeper than -3, but this rapidly softens to -5/2 as
high density structure emerges around 25\,Myrs. From t=30\,Myrs onwards, the
velocity power spectrum of the cloud complex has reached a steady state and is 
as shown in Fig. \ref{spectra}.

The velocity power spectra of Clump R is an order of magnitude less powerful
than that observed in the cloud complex as a whole. 
The cube encompassing Clump R for the Fourier Transform was 10\,pc on a side and
power spectra of Clump R are hence reduced by a factor of $4^3$ in order to allow
direct comparisons between cloud and Clump spectra.
Again the velocity power spectra has two inertial ranges over which the spectral indices 
are approximately constant, this time with a break between the two at $k=10$ - notably the
same physical scale of 1\,pc. At length-scales longer than 1\,pc (lower wavenumbers),
the inertial range is again -5/3 - indicating consistent sampling of the same thermal instability 
flow in the warm stable medium in this smaller box. At shorter length-scales, typically inside the clump, the index is
approximately -3, not dissimilar to that simulated by others on that scale and 
below \citep{medina14}.

It is remarkable that a simulation that started out as a stationary diffuse cloud and that
generates large-scale ordered flows, displays a long-lasting turbulence-like -5/3 spectrum
on an inertial range across one decade of wavenumber, 
{\it even though there is no fully developed turbulence in this simulation and there was no
driving scale in the initial condition}.
What the simulation has generated is flow containing a ``hierarchy of small-scale
irregularities superimposed on larger-scale more systematic motions'' -- very much the
definition of Larson-like turbulence 
\citep[][and ``1. Introduction'' therein for the source of this quote]{larson81}. This result
also makes it difficult to determine the nature of clumps, as what may appear turbulent
is actually a signature of complex infall velocities. Previous authors have noted that
clumps with clear signatures of infall are statistically indistinguishable from clumps with
no such signatures \citep{traficante18}. The same authors noted that the observed
non-thermal (i.e. supersonic) motions are not necessarily ascribed to turbulence acting
to sustain gravity, but they may be due to the gravitational collapse at clump scales -- 
precisely what we see here. 

Given the extensive nature of the warm neutral medium, this result suggests that this
index of -5/3 should extend to larger spatial scales than considered here - i.e. there 
is no physical reason for an inertial range limiting upper physical scale. It should
be noted that fully developed turbulence, even in astrophysical scenarios, displays
an inertial range over many orders of magnitude in $k$, for example in the magnetic 
fields of neutron stars \citep{wareing09,wareing10}. Galactic-scale simulations have 
found that the correlation scale of the large-scale random flows, calculated from the velocity
auto-correlation function, is on the order of 100\,pc \citep{gent13}. This may or may
not lead to a break in power spectra at this scale, but is larger than the scales considered
here, lending support to our finding of structured, or correlated, flows on scales shorter 
than 100\,pc. Further work is required to answer questions over how the large-scale 
turbulent nature of galaxies may be converted to correlated flows below the correlation
length. 

The lower physical scale limit of 
the turbulence-like inertial range is however a resolved physical characteristic of the 
models -- specifically, the scale of the cold condensations produced by thermal instability. It should 
not shift to smaller physical size (higher wavenumber) with the use of greater resolution.
Given the nature of the energy change across the phase boundary, with increasing 
infall velocity on the scale of a few km\,s$^{-1}$ on the warm side of the boundary,
rapidly decreasing to an order of magnitude lower velocities on the cold side of the
boundary, it would be natural to call the $\sim$few pc scale of these cold dense clumps 
the ``dissipative limit'' of the turbulent velocity spectrum of the warm neutral medium.

The power spectra of density in the cloud complex peaks at $k=8$. As noted 
previously, the size of the cube used for the Fourier Transform is 40\,pc on a side 
so this is equivalent to 5\,pc and in expected agreement with the 5\,pc scale of 
the dense clumps in the cloud. There is some sign of an inertial range
between $k=8$ and $k=40$ with an index around -5/3, reflective of the 
substructure in the clumps. Given there is structure 
throughout the cloud on the scale of 1 to 5\,pc as shown in Figs. \ref{wholecloud}
and \ref{coldens}, this seems reasonable. At $k>40$, there is a steepening 
of the power spectra, indicating relatively less smaller-scale structure. 
Calculation of density power spectrum at earlier times shows that before 23\,Myrs, the 
spectrum is far less powerful and weakly peaked at large $k$ on the small scale 
($\sim$1\,pc) of the growing inhomogeneities (as well as an expected strong peak
at the scale of the entire cloud complex). As structure grows in the simulation, the 
density power spectrum rapidly rises and steepens, although takes a longer period
to display a steady state spectral index than the velocity power spectrum. From 35\,Myrs onwards,
the density power spectrum is in agreement with the snapshot spectrum shown here in Fig.\ref{spectra}, 
although rising in power (shifting directly upwards in the figure) as densities increase in the clumps. 

Clump R displays a power spectra of density in agreement with this, with structure on the
order a few pc (up to $k=4$) before a steepening power spectra. If Clump R were
spherical with radius 5\,pc, we would expect a peak at $k=2$ from this 10\,pc box analysis, 
but as clearly shown in Fig. \ref{slices},
the clump is anything but spherical. There is no inertial range, but the spectral index
steepens from approximately -5/2, indicating that this clump, where self-gravity has
taken over from the effect of the thermal instability, is not dissimilar to those 
simulated elsewhere \citep{medina14}. There are also similarities with observations
of the Perseus molecular cloud, where multi-phase density power spectra with 
indices of -5/2 and steeper are derived \citep{pingel18}.

Again, it is worth highlighting that the power
spectra resulting from these simulations which start from a stationary initial condition,
are not dissimilar from those that start from turbulent initial conditions, nor dissimilar
to observed power spectra. There is also no evidence for the 0.29\,pc resolution
scale of the initial simulation in Paper IV  -- the spectra are smooth across 
$k$ of 138 for the cloud complex and $k$ of 34 for the Clump R spectra).
If anything can be considered a `driving scale' in these simulations, it is that 0.29\,pc
scale upon which the initial density inhomogeneities are introduced at $t=0$. There is
no evidence that this scale has dominated our simulations and it is very different
to the 5\,pc acoustic length scale of the warm stable medium that seems to be
observed defining clump-scales here and sheet separation scales in the magnetic
case (see Papers I and IV). 

Further study is required to determine whether the
use of power spectra can meaningfully discern between models of star formation,
but these results indicate power spectra are a blunt tool for understanding the
formation of molecular clouds. The results also cast doubt on any justification for
the injection of driven turbulence in numerical models of molecular clouds, since it
is apparent that our simulations naturally produce long-lasting turbulence-like 
power spectra from an ordered flow. The  question of whether such flows  can reproduce other 
observational features  of the velocity and density dispersions, is left to a later paper.

\section{Re-simulating the final collapse of the most massive clump}\label{collapse}

The most massive clump, Clump D, has been extracted and resimulated at even higher resolution 
(0.016\,pc on the finest AMR level) in order to examine the collapse of the clump and confirm that
it is gravitationally bound and Jeans unstable. On a short time scale of approximately 100,000
years, from t=44\,Myrs to t=44.1\,Myrs, the density in the clump rises steeply up to 
n$_{\rm H}=3.5\times10^6$\,cm$^{-3}$. 91\% of the mass in the clump, equivalent to
approximately 250\,M$_{\odot}$, is in the cold phase. The unstable envelope around the
clump contains little more than one solar mass of material. The central pressure in the clump
has risen to two orders of magnitude greater than the average pressure of the warm stable
surroundings. An energy analysis of the simulated data reveals that the clump is
strongly gravitationally bound - the clump has three times more gravitational energy than
the sum of kinetic and thermal energy, making it also Jeans unstable. The apparent nature
of the clump is very similar to that shown in Fig. \ref{slices}, albeit with a far greater peak
density. A Plummer-like profile is a good fit, but again substructure in the clump means there
are large variations in different radial directions outward from the peak density. 

In combination with the simulations presented earlier, we have conclusively demonstrated that
thermally unstable diffuse material can evolve cold and dense structure which can be strongly
influenced by gravity. This can lead to the eventual collapse of such structure under gravity, 
without the need to include any other physics or factors, e.g. pressure waves, shocks
or collisions, although clearly these play a role which can now be elucidated in future work.
Remarkably, in the purely hydrodynamic case, the natural evolution of the diffuse unstable
cloud produces a distribution of cold, dense clumps connected by long-lived filaments that
would appear to have a characteristic size scale of 0.2 to 0.6 pc.

\section{Conclusions}\label{conclusions}

In this work, we set out to determine by the use of hydrodynamic simulation whether 
a molecular cloud structure evolved from diffuse, thermally unstable medium could ever 
lead to gravitationally collapsing structure, without the influence of any other physics
(e.g. turbulence) or external disturbance (pressure wave, shock or collision).

Our previous work, at lower resolution of 0.29\,pc or greater, had revealed that clumpy clouds form in the
hydrodynamic case and corrugated sheet-like clouds, that in projection appear
filamentary, form in the magnetic case. Neither set of previous simulations conclusively
demonstrated gravitational collapse within such models of molecular clouds. 

The suite of high-resolution simulations carried out here, with typically 0.078\,pc resolution,
but up to 0.016\,pc resolution, have now conclusively demonstrated that thermal
instability in a diffuse medium alone can generate cold and dense enough structure to
allow self-gravity to take over and conclude the star formation process. The total
time scale for this to happen is on the order of 40\,Myrs, although the structure would
only be considered a molecular cloud for the previous 15\,Myrs. The final gravitational
collapse sees density increase by 3 orders of magnitude on very short time scales of
10$^5$ years.

We have noted the following:-
\begin{enumerate}
\item Diffuse thermally unstable material evolves into a stable network of cold dense
clumps multiply-connected by filaments, immersed in warm tenuous material. 
\item The filaments form and accrete as unstable gas falls in, with widths
ranging from less than 0.26\,pc as they first form, up to 0.56\,pc and more after
6\,Myrs.
\item During the early evolution of the cloud complex, high densities in the 
filaments and proto-clumps caused by
thermal instability driven flow can be mis-leading, in terms of applying automatic
star particle injection routines. Convergent flow, energy analysis and gravitational
potential conditions can all be satisfied, but accurate investigation reveals this
to be a transient dynamical phase in the formation of the cloud complex, not
gravitational collapse.
\item Application of the FellWalker \citep{berry15} algorithm identifies 21
massive ($>20$\,M$_{\odot}$) clumps that have formed in this 3,000\,M$_{\odot}$
region of cloud.
\item The clumps formed have a size-scale of 5\,pc or less, masses up to 
300\,M$_{\odot}$, internal temperatures of 10-30\,K and internal
velocity dispersions of 0.5\,km\,s$^{-1}$ or less. These physical characteristics are in agreement with the definition
of clumps presented by \cite{bergin07} and density distributions can be fitted
by a Plummer-like profile \citep{whitworth01}.
\item In agreement with previous models \citep[see][and references therein]{micic13} 
the majority of clumps are not initially collapsing. The most massive clump has 
been investigated at a resolution of 0.016\,pc and does collapse under gravity, increasing
its central density by 3 orders of magnitude on a time scale of 100,000 years.
\item Velocity power spectra of the cloud complex as a whole and an individual 
massive clump show spectral indices which are turbulence-like (-5/3) over a short
inertial range (approximately one decade of wavenumber), even though the initial
diffuse condition was stationary. This is the result of a large-scale flow with
a hierarchy of small-scale structure, very much as predicted by \cite{larson81}.
The wavenumber of the break-point in the spectrum corresponds to the 5\,pc 
size-scale of individual clumps.
\item Power spectra of velocity and density are not dissimilar to simulations that
employ turbulent initial conditions implying that 1D power spectra may not offer a 
meaningful tool to discern between models of star formation.
\item The most massive clumps eventually undergo runaway gravitational collapse
with analysis determining that they are truly gravitationally bound and Jeans
unstable.
\end{enumerate}

Immediate future work will consider the effect of magnetic fields in this scenario,
applying the same high resolution to MHD simulations (which have been run along
side this suite of simulations) and looking for gravitational collapse in the magnetic
case, as well as explanations for the origins of striations \citep{tritsis16},
disconnections along filaments resulting in `integral-shapes' \citep{stutz18}
and the formation of strongly distorted magnetic fields, e.g. hour-glass field 
morphologies \citep{pattle17}. 
Other work will study the effect of various feedback mechanisms in this cloud complex, 
introducing a robust star-particle formation technique, sampling a realistic initial mass 
function in order to determine the star formation rate of this cloud complex.

\section*{Acknowledgments}

We acknowledge support from the Science and Technology Facilities
Council (STFC, Research Grant ST/P00041X/1). The calculations 
herein were performed on
the DiRAC 1 Facility at Leeds jointly funded by STFC, the Large
Facilities Capital Fund of BIS and the University of Leeds and on
other facilities at the University of Leeds. 
Data presented herein is available from http://doi.org/10.5518/483 .
We thank the anonymous referee for a detailed review of the manuscript
which improved the presentation and flow of the manuscript considerably,
and which enhanced the detail of a number of the results presented.
We thank David Hughes at Leeds for the provision of IDL scripts which
formed the basis of the power spectra analysis presented in this work.
VisIt \citep{visit} is supported by the Department of Energy with funding
from the Advanced Simulation and Computing Program and the
Scientific Discovery through Advanced Computing Program.

\label{lastpage}

\end{document}